\documentclass[conference]{IEEEtran}
\IEEEoverridecommandlockouts
\usepackage{cite}
\usepackage{amsmath,amssymb,amsfonts}
\usepackage{algorithm, algorithmic}
\usepackage{graphicx}
\usepackage{textcomp}
\usepackage{xcolor}
\usepackage{bm}
\usepackage{url}
\usepackage[hidelinks]{hyperref}
\usepackage{subcaption}
\def\BibTeX{{\rm B\kern-.05em{\sc i\kern-.025em b}\kern-.08em
    T\kern-.1667em\lower.7ex\hbox{E}\kern-.125emX}}
\usepackage{flushend}

\usepackage{enumitem}

\newtheorem{theorem}{Theorem}[section]

\begin{document}

\makeatletter
    \newcommand{\linebreakand}{%
      \end{@IEEEauthorhalign}
      \hfill\mbox{}\par
      \mbox{}\hfill\begin{@IEEEauthorhalign}
    }
\makeatother

\title{A Differentially Private Blockchain-Based Approach for Vertical Federated Learning}

\author{\IEEEauthorblockN{Linh Tran}
\IEEEauthorblockA{\textit{Dept. of Computer Science} \\
\textit{Rensselaer Polytechnic Institute}\\
Troy, NY \\
tranl3@rpi.edu}
\and
\IEEEauthorblockN{Sanjay Chari}
\IEEEauthorblockA{\textit{Dept. of Computer Science} \\
\textit{Rensselaer Polytechnic Institute}\\
Troy, NY \\
charis3@rpi.edu}
\and
\IEEEauthorblockN{Md. Saikat Islam Khan}
\IEEEauthorblockA{\textit{Dept. of Computer Science} \\
\textit{Rensselaer Polytechnic Institute}\\
Troy, NY \\
islamm9@rpi.edu}
\linebreakand
\IEEEauthorblockN{Aaron Zachariah}
\IEEEauthorblockA{\textit{Dept. of Computer Science} \\
\textit{Rensselaer Polytechnic Institute}\\
Troy, NY \\
zachaa2@rpi.edu}
\and
\IEEEauthorblockN{Stacy Patterson}
\IEEEauthorblockA{\textit{Dept. of Computer Science} \\
\textit{Rensselaer Polytechnic Institute}\\
Troy, NY \\
sep@cs.rpi.edu}
\and
\IEEEauthorblockN{Oshani Seneviratne}
\IEEEauthorblockA{\textit{Dept. of Computer Science} \\
\textit{Rensselaer Polytechnic Institute}\\
Troy, NY \\
senevo@rpi.edu}
}

\maketitle

\begin{abstract}
We present the Differentially Private Blockchain-Based Vertical Federal Learning (DP-BBVFL) algorithm that provides verifiability and privacy guarantees for decentralized applications. DP-BBVFL uses a smart contract to aggregate the feature representations, i.e., the embeddings, from clients transparently. We apply local differential privacy to provide privacy for embeddings stored on a blockchain, hence protecting the original data.
We provide the first prototype application of differential privacy with blockchain for vertical federated learning. Our experiments with medical data show that DP-BBVFL achieves high accuracy with a tradeoff in training time due to on-chain aggregation. 
This innovative fusion of differential privacy and blockchain technology in DP-BBVFL could herald a new era of collaborative and trustworthy machine learning applications across several decentralized application domains.
\end{abstract}

\section{Introduction}
Federated Learning (FL) \cite{mcmahan2023communicationefficient} is a form of machine learning devised to allow learning from data without compromising its confidentiality. In FL, multiple clients train a machine learning model on data local to the client and share only the intermediate information, for example, model updates, with a central server.
FL has applications in multiple sectors like healthcare, financial services, and supply chain management~\cite{Wang2022}.
Horizontal Federated Learning (HFL)~\cite{yang2019federated} is a form of FL where clients share the same feature set, but each has a different set of samples.
Vertical Federated Learning (VFL)~\cite{10.1145/3292500.3330765, 9463409} is another form of FL where clients have the same set of samples, but each has a different feature set.

While FL is designed to keep the client's raw data locally, it has been shown that the shared intermediate information can leak the original private data \cite{geiping2020inverting, mahendran2014understanding}.
Against this backdrop, it is worth noting that the majority of existing privacy-preserving solutions have concentrated on HFL, where different clients possess different data samples of the same feature set, leaving a significant gap in privacy protections for VFL scenarios. Given the diverse applications of VFL, ranging from healthcare to finance, where the confidentiality of shared information is paramount, addressing this gap is not just a technical challenge but a necessity for ethical and secure data utilization.

While VFL with blockchain provides transparency and verifiability for the aggregation process and client data protection, it does not address the issue of client raw data leaked from the embeddings. Since the embeddings stored on a blockchain are widely accessible, safeguarding the embeddings is essential in order to protect the raw data. An approach commonly used to address this issue is Differential Privacy (DP)~\cite{10.1561/0400000042}. DP is a method that reduces the probability of data extraction by adding calibrated noise to the data.
Local DP is a subcategory of DP that injects privacy noise locally at each client site before any data sharing. There have been multiple works that use DP to strengthen the privacy of data in both HFL~\cite{wei2019federated, truex2020ldpfed, truex2019hybrid} and VFL~\cite{tran2023privacy}. 
However, our solution stands at the forefront of innovation by being the first to merge DP and blockchain technologies with VFL to provide a privacy guarantee and verifiability for decentralized applications.


\subsection{Motivating Scenario}
In a healthcare scenario, Hospital A might have diagnostic information (blood tests, X-rays, CT scans, MRI results) for a set of patients, while Hospital B has lifestyle data (diet, exercise routines) for the same patients. Here, the data is vertically partitioned because both hospitals have information about the same patients (overlapping rows) but different kinds of information (non-overlapping columns).
Collectively, all these data are valuable for training models to assist in diagnosing diseases for a common set of patients. For example, health data from global pandemics such as COVID-19 could be used to make critical insights like outbreak prediction that could save millions of lives. However, hospitals are typically bound by patient privacy regulations, like the Health Insurance Portability and Accountability Act (HIPAA) in the United States or European Health Data Space (EHDS) regulation in Europe, that prevent patients' medical data from being shared directly.
In this context, VFL could be leveraged to gather insights from each hospital without compromising patient data privacy.
In such a VFL algorithm, each client has a local network that transforms their raw features into embeddings. These embeddings are aggregated using summation, and a central server uses the embedding sum to train a \emph{fusion model} that produces the predicted labels. Clients involved in a VFL system need to place an implicit level of trust that the organization managing the server will behave fairly and without bias since they cannot explicitly verify the operations performed on the server. Especially for healthcare applications, this could be a deciding factor if each party (such as a hospital) wants to participate and contribute their private medical data and does not have 100\% trust in the organization that is hosting the server. 
Blockchain provides a transparent method to compute and store embeddings and transaction details that clients can access to verify the fairness of the computation.
However, by doing so, blockchain reveals the embedding values to all participants, leading to a data privacy leak. In this case, DP comes in as a solution to protect the private data that is publicly stored on blockchain.


\subsection{Contributions}
We propose DP-BBVFL, a \emph{serverless} VFL algorithm designed for different institutions with vertically distributed data to collaborate and train a machine learning task that leverages blockchain for aggregation of the \texttt{embeddings} to provide transparency and verifiability to the training process, and a local DP mechanism to ensure data privacy for participating clients.
Our main contributions are as follows:

\begin{itemize}[noitemsep,topsep=0pt,itemsep=0pt]
    \item Differentially private VFL system that does not need a trusted third party.
    \item Verifiable and transparent training for institutions using blockchain technology for aggregation.
    \item Provable privacy bound of the algorithm throughout the training process with detailed analysis.
    \item Demonstration of the DP-BBVFL algorithm that provides high accuracy on the Breast Cancer~\cite{misc_breast_cancer} and MIMIC-III~\cite{mimic-article} datasets while providing verifiability and DP guarantee.
\end{itemize}


The DP-BBVFL method uniquely combines DP and blockchain within a VFL framework, offering both enhanced privacy and verifiability compared to other methods. Further, this innovative approach eliminates the need for a central server, reducing vulnerability points and decentralizing data control, which is critical in sectors like healthcare and finance, where data integrity and privacy are paramount.
The contribution of DP-BBVFL is significant as it sets a new standard for secure and trustworthy machine-learning applications in decentralized environments.

\section{Related work} 

In recent years, VFL has become a key strategy for collaborative machine learning where multiple clients want to train their vertically distributed data without sharing the raw data. However, there is possible data leakage from the shared embeddings. Numerous works have addressed this issue by applying DP. For example, Das et al.~\cite{das2022cross} and Hu et al.~\cite{10.1145/3292500.3330765} have explored FL algorithms that incorporate DP for improving privacy and communication efficiency by maintaining the data partitioning across different silos. 
Tran et al.~\cite{tran2023privacy} and Tang et al.~\cite{Tang2023} introduce novel frameworks for feature privacy and multi-party computation to secure data during the training process. In addition, Adnan et al.~\cite{adnan2022federated} emphasizes the importance of DP in medical image analysis. Although all these differentially private approaches provide privacy protection in VFL, they often lack transparency and verifiability to ensure the integrity of the aggregation process. Our work addresses this issue by integrating blockchain technology with VFL while offering a decentralized solution that reduces reliance on a central server.

The integration of blockchain technology with FL has led to the emergence of BlockChain-based Federated Learning (BCFL). BCFL introduces novel solutions to address the FL scalability challenges and also provide verifiability. For instance, Wang et al.~\cite{wang2021blockchainbased}, and Drungilas et al.~\cite{drungilas2021towards} elucidate how blockchain can mitigate scalability hurdles while combining with HFL, and Unal et al.~\cite{unal2021integration} combine HFL and blockchain to prevent model poisoning attacks.
Other BCFL methods that integrate blockchain within VFL systems include ~\cite{9987097, cui2022fast, cmes.2023.026920} and \cite{Tang2023}.
Most of these works did not use DP within the BCFL framework to protect the data, which might have a detrimental impact on domains like healthcare and finance.
The only exception is the work by Tang et al.~\cite{Tang2023} that proposed an intention-hiding BCFL using DP. However, they only focus on protecting the model training intention, and our approach aims to provide privacy for the whole private training data set.
We utilize DP in our work, which provides strong and provable privacy in VFL and prevents the leakage of private medical and financial data.
Our approach further extends the applicability of BCFL beyond the healthcare and financial domain, which provides a new paradigm in privacy-preserving distributed learning.

\section{Background}

In this section, we present background information on DP, PBM, and blockchain, along with the pertinent components incorporated into our algorithm to maintain privacy in VFL.
 
\subsection{Differential Privacy}

Differential Privacy~\cite{10.1561/0400000042} provides a mathematical guarantee of privacy by allowing data to be analyzed without revealing sensitive information about any individual in the dataset. This is achieved by adding random noise to the data to make it impossible to infer any specific individual entries used for training.
In this work, we employ R\'{e}nyi Differential Privacy (RDP) \cite{Mironov_2017}, a specialized version of DP that is derived from the principle of the R\'{e}nyi divergence \cite{van2014renyi}. We give a formal definition of R\'{e}nyi divergence and RDP as follows.

\textbf{Definition 3.1:} For $\alpha \in (0, \infty), \alpha \neq 1$, the \textit{R\'{e}nyi divergence} of order $\alpha$ between two probability distributions $P$ and $Q$ over a set $X$ is:
\[
D_{\alpha}(P \| Q) = \frac{1}{\alpha - 1} \log \sum_{x \in X} P(x)^{\alpha} Q(x)^{1-\alpha}.
\]

\textbf{Definition 3.2:} A randomized mechanism $M$ satisfies ($\alpha, \epsilon$)-\textit{RDP} if for any two adjacent datasets $D$ and $D'$ and for any $\alpha > 1$, it holds that:
\[
D_{\alpha}(M(D) \| M(D')) \leq \varepsilon.
\]

We utilize RDP because it offers more nuanced control over the level of privacy or the privacy budget, allowing for finer adjustments in privacy-accuracy trade-offs compared to standard DP. In addition, RDP simplifies the composition analysis, providing tighter bounds on the cumulative privacy loss for learning algorithms with multiple training rounds. RDP's framework, based on R\'{e}nyi divergence, enhances analytical clarity, making it easier to understand and quantify the privacy guarantees of various mechanisms.

\subsection{Poisson Binomial Mechanism}
A crucial element of our algorithm involves ensuring the privacy of the client's raw data by protecting the embeddings during aggregated sum computation. To achieve this goal, we rely on the Poisson Binomial Mechanism (PBM) proposed by Chen et al.~\cite{chen2022poisson} for HFL. Tran et al.~\cite{tran2023privacy} extend the PBM technique for VFL settings with detailed privacy analysis and tradeoffs for the whole training duration.
We utilize PBM because it is a privacy-preserving method that provides an RDP guarantee with unbiased aggregated value. It is also one of the very few methods that have been shown and proven to be applicable in VFL systems.
At its core, the PBM provides an RDP guarantee by injecting controlled binomial noise into the embeddings before the aggregation process. 
Simultaneously, the PBM quantizes the embeddings with the binomial distribution, which makes the noisy embeddings output quantized integer values. 
This quantization method aids blockchain integration as smart contracts often handle integer values more efficiently than floating-point numbers due to the deterministic nature of blockchain computations and the limitations in handling non-integer arithmetic.

The Scalar PBM provided in Algorithm~\ref{alg:scalar_poisson_binomial} is the process to compute a summation using PBM with a certain RDP guarantee.
Suppose there are $M$ clients, each has private input \(a_i \in [-C, C]\), and they want to compute the sum \(s = \sum_{i=1}^{M} a_i\). Each client applies Algorithm~\ref{alg:scalar_poisson_binomial} to their respective value and gets their quantized output $q_i$. The noisy quantized sum is computed as \(\hat{q} = \sum_{i=1}^{M} q_i\). An estimation value of \(s\) is then calculated as \(\frac{C}{\beta b} (\hat{q} - \frac{bM}{2})\).
This sum computation yields an unbiased estimated sum with variance $\frac{C^2M}{4\beta^2 b}$, while providing $(\alpha, \epsilon(\alpha))$-RDP guarantee for any $\alpha > 1$ and $\epsilon(\alpha) = \Omega (b\beta^2 \alpha)$, following the work by Tran et al.~\cite{tran2023privacy}.

\begin{algorithm}[t]
    \caption{Scalar Poisson Binomial Mechanism}
    \label{alg:scalar_poisson_binomial}
    \begin{algorithmic}[1]
    \STATE {\textbf{Initialize:}} $a_i \in [-C, C]$; $b \in \mathbb{N}$, $\beta \in [0, \frac{1}{4}]$.
    \STATE Compute probability $p_i \leftarrow \frac{1}{2} + \frac{\beta}{C}a_i$.
    \STATE Generate quantized value $q_i \leftarrow \text{Binom}(b, p_i)$.
    \STATE \textbf{Output:} Quantized value $q_i$.
    \end{algorithmic}
\end{algorithm}

\subsection{Blockchain}
We propose replacing the central server leveraged in conventional HFL with a blockchain component as the aggregator to receive the client's embeddings and perform the sum aggregation. By using blockchain, we can address two problems - the need for trust and the need for visibility. 
Blockchain's inherent characteristics, such as immutability and consensus mechanisms, ensure that once data or transactions (in this case, embeddings and aggregation results) are transparently recorded, they cannot be altered without consensus.
This also means that the aggregation process and the resulting outputs are open for inspection, promoting transparency and allowing participants to verify that their data has been accurately and fairly included in the model training process.
Additionally, since the aggregator smart contract is not beholden to any central authority but resides on a private permissionless blockchain network, clients need not trust other clients or the aggregator. We make use of a private permissionless blockchain because having an entity that governs the blockchain would reduce the transparency of our solution.

Furthermore, we implement a simple reward system using fungible tokens, which is implemented to be compliant with Ethereum's ERC20 standard~\cite{ethereumERC20Token}.
These incentives can encourage participation and contribution to the VFL process, enhancing the model's overall performance and data diversity.
Organizations or individuals interested in accessing the insights or models developed through the VFL process might use fungible tokens to pay for access. This could eventually create a marketplace for models and insights where tokens serve as a medium of exchange, facilitating transactions between data consumers and the DP-BBVFL ecosystem.


\section{Methodology}

\begin{figure}[h]
\centering
\includegraphics[scale=0.25]{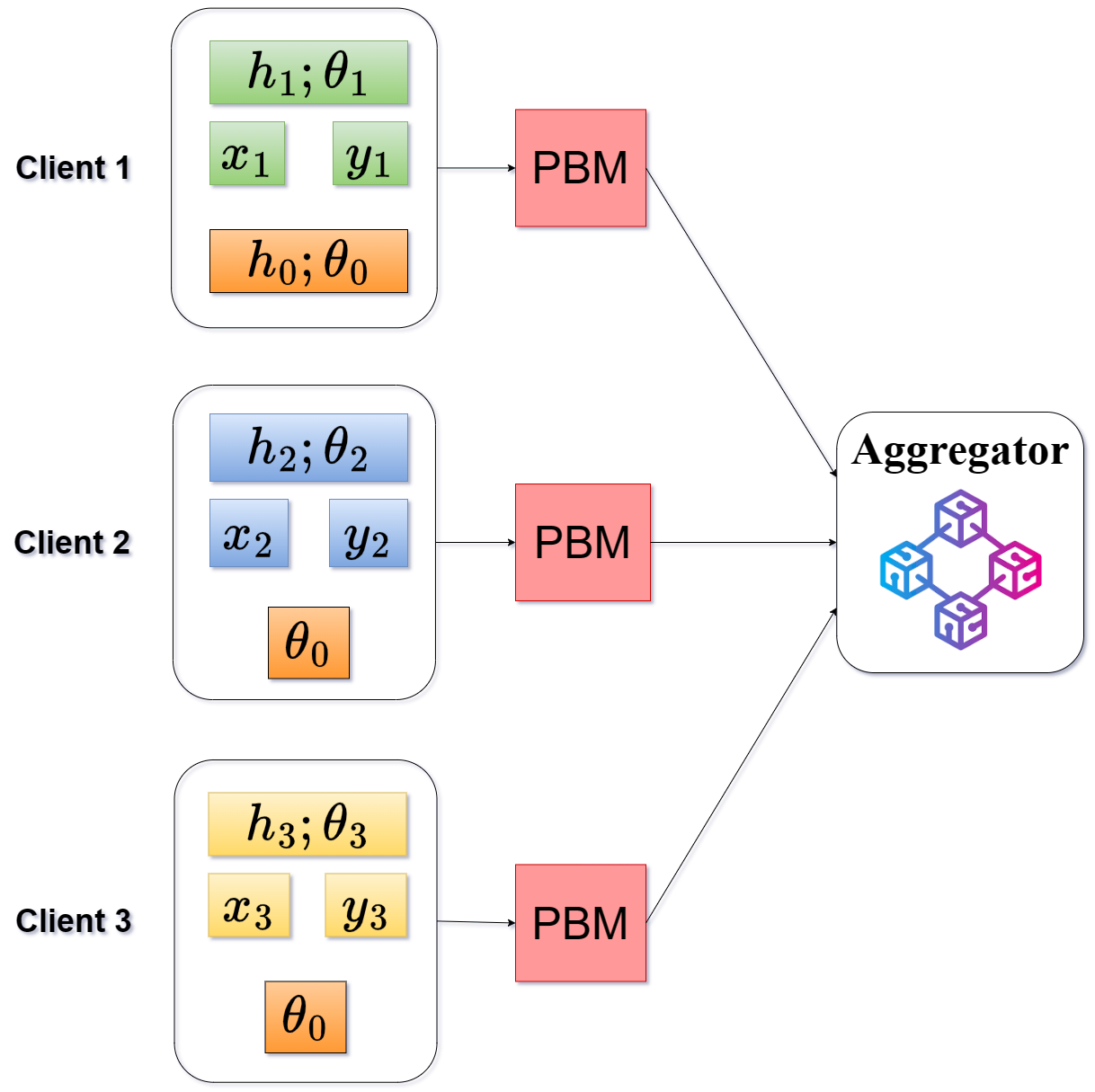}
\caption{Example of system model with $3$ clients.}
\label{NNmodel}
\vspace{-3ex}
\end{figure}

\subsection{System Model}

The joint training dataset $\textbf{X} \in \mathbb{R}^{N \times D}$ is partitioned across the $M$ clients, where $N$ is the number of data samples and $D$ is the number of features. 
Let $\textbf{x}^i$ denote the $i$th sample of $\textbf{X}$. 
For each sample $\textbf{x}^i$, each client $m$ holds a disjoint subset $\textbf{x}_m^i$ of the features, where $\textbf{x}^i = [\textbf{x}_1^i \ldots \textbf{x}_M^i]$.
The entire vertical partition of $\textbf{X}$ that is held by client $m$ is denoted by $\textbf{X}_m$, where $\textbf{X} = [\textbf{X}_1 \ldots \textbf{X}_M]$.
Let the set of all labels be $\textbf{y}\in \mathbb{R}^{N \times 1}$, and $\textbf{y}_i$ be the label of sample $i$. Each client $m$ holds a disjoint subset $\textbf{y}_m$ of $\textbf{y}$. Clients do not need to share their labels during training and can compute the loss function for samples with labels they hold locally.

The overall ``system'' model consists of $M$ clients who want to train a global model collaboratively and solve a common task over their vertically distributed data.
The global model includes $M$ clients' local neural network models and a fusion model. The fusion model is a neural network model that takes aggregated outputs of $M$ client local models and predicts the corresponding labels.
Before any training process, one client is selected as the active client responsible for training and updating the fusion model, while the other clients only have access to the fusion model parameters.
The active client can be selected randomly using verifiable random functions or by using a democratic mechanism implemented using a smart contract based governance mechanism. Since the active client monitors the fusion model, there may be a need for trust in the active client selection; however, protecting the fusion model is not part of DP-BBVFL goals. Therefore a single active client was used in our experimentation.

Each client $m$'s local neural network is denoted as $h_m$ with parameters $\bm\theta_m$. Each client $m$ uses $h_m$ that takes their private data $\textbf{X}_m$ as input and produces an embedding.
All clients' embeddings are summed, and this sum is the input for the fusion model $h_0$.
Computing the embedding sum is done by a blockchain-based aggregator, which is a Solidity smart contract that provides an incentive in the form of an ERC-20 token to all clients.
An example of the system model with $3$ clients is shown in Fig.~\ref{NNmodel}.

\subsection{Algorithm}

\begin{algorithm}[t]
    \caption{DP-BBVFL}
    \label{alg}
    \begin{algorithmic}[1]
    \STATE {\textbf{Initialize:}} $\bm\theta_m^0$, $m=0 \ldots M$; $b>0, 0<\beta<\frac{1}{4}$.
    \FOR {$t \leftarrow 0 \ldots T-1$}
        \STATE Randomly sample $\mathcal{B}^t$ from $(\textbf{X}, \textbf{y})$
        \FOR {$m \leftarrow 1 \ldots M$ in parallel}
            \STATE  $q_m^t \leftarrow \textbf{PBM}(h_m(\bm\theta_m^t ; \textbf{X}^{\mathcal{B}^t}_m), b, \beta)$
            \STATE Client $m$ sends $q_m^t$ to the aggregator
        \ENDFOR
        \STATE Aggregator computes $\hat{q}^t \gets \sum_{m=1}^M q_m^t$
        \STATE Aggregator sends $\hat{q}^t$ to all clients
        \STATE All clients compute $\hat{h}^t \leftarrow \frac{1}{\beta b} (\hat{q}^t - \frac{bM}{2})$
        \FOR {$(\textbf{X}^i, \textbf{y}^i) \in \mathcal{B}^t$}
            \STATE Client $m$, $\textbf{y}^i \in \textbf{y}_m$ computes and sends
            \STATE $~~~~\nabla_{\hat{h}} \ell\left(\bm \theta_0, \hat{h}^t; \textbf{y}^i \right)$ to active client
        \ENDFOR
        \STATE /* At active client */
        \STATE $\nabla_{\hat{h}} \mathcal{L}_{\mathcal{B}^t}(\bm\theta_0^t, \hat{h}^t) \leftarrow \frac{1}{B} \sum_{(\textbf{X}^i, \textbf{y}^i) \in \mathcal{B}^t} \nabla_{\hat{h}} \ell\left(\bm \theta_0, \hat{h}^t; \textbf{y}^i \right)$
        \STATE Active client sends $\nabla_{\hat{h}} \mathcal{L}_{\mathcal{B}^t}(\bm\theta_0^t, \hat{h}^t)$ to all clients
        \STATE $\bm\theta_0^{t+1} \gets \bm\theta_0^t - \eta \nabla_0 \mathcal{L}_{\mathcal{B}^t}(\bm\theta_0^t, \hat{h}^{t})$
        \FOR {$m \leftarrow 1 \ldots M$ in parallel}
            \STATE $\nabla_m \mathcal{L}_{\mathcal{B}^t} \gets \nabla_m h_m(\bm\theta_m^t; \textbf{X}_m^{\mathcal{B}^t}) \nabla_{h_m} \hat{h}^{t}  \nabla_{\hat{h}} \mathcal{L}_{\mathcal{B}^t}(\bm\theta_0^t, \hat{h}^t)$
            \STATE $\bm\theta_m^{t+1} \gets \bm\theta_m^t - \eta^t \nabla_m \mathcal{L}_{\mathcal{B}^t}$
        \ENDFOR
    \ENDFOR
\end{algorithmic}
\end{algorithm}

The complete DP-BBVFL algorithm is given in Algorithm~\ref{alg}.
At the initial stage, each client $m$ sets up their starting local parameters $\bm\theta_m^0$, $m=1 \ldots M$, and the active client sets up the fusion model parameter $\bm\theta_0$ (line $1$). Since all clients hold a disjoint set of labels needed for backpropagation, the parameter $\bm\theta_0$ is shared among clients throughout the training process. The PBM parameters $b, \beta$ are chosen to achieve a certain level of privacy. As part of the data pre-processing step, all clients need to align the shared sample IDs without revealing their private feature data. This step can be done using Private Entity Resolution \cite{xu2021fedv}.

The algorithm runs for $T$ training iterations. In each iteration $t$, all clients decide on a minibatch $\mathcal{B}^t$ randomly drawn from $\textbf{X}$ (line $3$).
The clients do this by each using a pseudorandom generator initialized with the same seed. Thus, each client picks the same list of sample IDs for the minibatch.

From line $4$ to line $7$, each client $m$ uses the local network $h_m$ with parameters $\bm\theta_m^t$ that takes each vertically partitioned sample $\textbf{x}^i_m$, $i \in \mathcal{B}^t$ as input and produces an embedding $h_m(\bm\theta_m^t; \textbf{x}^i_m)$ with dimension $P$. We denote the embedding set of minibatch $\mathcal{B}$ produced by client $m$ as $h_m(\bm\theta_m^t ; \textbf{X}^{\mathcal{B}^t}_m)$.
Each client $m$ applies PBM with parameters $b, \beta$ on their embedding set $h_m(\bm\theta_m^t ; \textbf{X}^{\mathcal{B}^t}_m)$, and gets the noisy quantized embedding set $q_m^t$.
Each client $m$ then sends $q_m^t$ to the aggregator smart contract for aggregation.

The aggregator smart contract computes the summation $\hat{q}^t$ and sends the result to all clients (lines $8-9$). In addition, the aggregator sends a token to the client's address, which quantifies the client's contribution to the training process.
All clients compute the estimated embedding sum $\hat{h}^t = \frac{1}{\beta b} (\hat{q}^t - \frac{bM}{2})$ (line $10$). This sum is needed to calculate the gradient of the loss function over $\mathcal{B}^t$, which is denoted by $\nabla_{\hat{h}} \mathcal{L}_{\mathcal{B}^t}$.

We note that $\textbf{y}^{\mathcal{B}^t}$ may be distributed among clients. Therefore, to compute $\nabla_{\hat{h}} \mathcal{L}_{\mathcal{B}^t}$, all clients with $| \textbf{y}_m \cap \textbf{y}^{\mathcal{B}^t} | > 0$ take turn to compute $\ell (\bm\theta_0, \hat{h}; \textbf{y}_i)$ of each sample $(\textbf{x}^i,\textbf{y}^i) \in \mathcal{B}$ individually.
Each client sends their computed gradients to the active client (lines $11$-$14$). The active client then computes the average gradient $\nabla_{\hat{h}} \mathcal{L}_{\mathcal{B}^t}$ and sends the result to all other clients (lines $16$-$17$). The average gradient $\nabla_{\hat{h}} \mathcal{L}_{\mathcal{B}^t}$ is computed as follows:

\begin{align}
    \nabla_{\hat{h}} \mathcal{L}_{\mathcal{B}^t} \gets \frac{1}{B} \sum_{(\textbf{x}^i,\textbf{y}^i) \in \mathcal{B}} \nabla_{\hat{h}} \ell\left(\bm\theta_0, \hat{h}(\bm \theta_1 \ldots \bm \theta_M; \textbf{x}^i); \textbf{y}^i \right).
\end{align}

Finally, the active client updates the fusion model parameters for the next iteration $\bm\theta_0^{t+1}$ (line $18$). Each client uses the average gradient $\nabla_{\hat{h}} \mathcal{L}_{\mathcal{B}^t}$ to compute the partial derivative of $\mathcal{L}_{\mathcal{B}^t}$ with respect to its local model parameters $\bm\theta_m^t$ using the chain rule as:

\begin{align}
    \nabla_m \mathcal{L}_{\mathcal{B}^t} \gets \nabla_m h_m(\bm\theta_m^t; \textbf{X}_m^{\mathcal{B}^t}) \nabla_{h_m} \hat{h}^{t}  \nabla_{\hat{h}} \mathcal{L}_{\mathcal{B}^t}(\bm\theta_0^t, \hat{h}^t),
\end{align}

where $\nabla_{h_m} \hat{h}^{t}$ is the identity operator. Each client then updates its local model parameters for the next iteration $\bm\theta_m^{t+1}$ (lines $19$-$22$).

\subsection{Privacy Analysis}

By the definition of RDP, it is guaranteed that Algorithm~\ref{alg} protects the input data, i.e., the feature sets of all clients, with indistinguishability. In other words, no one can recover the features used for a particular observed training output with high probability. The probability to which an attacker can recover the input features, or the privacy budget, depends on the privacy parameter $\epsilon$. Since the PBM is used on randomly selected data repeatedly throughout the training process, there is a chance that some data is observed multiple times, which aids the attacker in recovering the private input features, increasing the privacy budget.
We give an accounting of the privacy budget across $T$ training iterations of Algorithm~\ref{alg}.

\begin{theorem}  \label{privacybudget.thm} 
For any $b\in \mathbb{N}$ and $\beta \in [0, \frac{1}{4}]$, Algorithm~\ref{alg} satisfies $(\alpha, \epsilon(\alpha))$-RDP for any $\alpha > 1$ and where $C_0$ is a universal constant.

\begin{align}
    \epsilon(\alpha) = C_0 \frac{TBPb\beta^2 \alpha}{N},
\end{align}

\end{theorem}

\noindent\textbf{Proof:} Following from Tran et al.~\cite{tran2023privacy}, a single iteration of Algorithm~\ref{alg} satisfies RDP with $\epsilon(\alpha) = C_0 Pb\beta^2 \alpha$, when publicizing each individual noisy embeddings of size $P$ for each sample on-chain. For the full dataset of each client, the privacy budget remains the same with $\epsilon(\alpha) = C_0 Pb\beta^2 \alpha$ by Parallel Composition \cite{10.1561/0400000042}. At each iteration, each sample is chosen for training with a rate of $B/N$. Sequentially, Algorithm~\ref{alg} across all $T$ iterations has a privacy budget of $\epsilon(\alpha) = C_0 \frac{TBPb\beta^2 \alpha}{N}$.

We note that lowering the value of $\epsilon$ gives a higher level of privacy protection. However, this generally lowers the utility of the learning algorithm as a tradeoff. Theorem~\ref{privacybudget.thm} estimates the privacy budget $\epsilon(\alpha)$ of Algorithm~\ref{alg}, and we can use this to balance the privacy-utility tradeoff. One can increase the level of privacy guarantee of Algorithm~\ref{alg} by decreasing the PBM noise parameters $b$ or $\beta$. Other machine learning parameters, including number of iterations $T$, batch size $B$, embedding size $P$, and number of features $N$, also affect the overall privacy budget of Algorithm~\ref{alg}. However, these parameters are typically chosen to optimize the utility, and they are considered fixed terms when computing the privacy budget.

\begin{figure}[t]
\centering
\begin{subfigure}[t]{0.24\textwidth}
\centering
  \includegraphics[width=\textwidth] {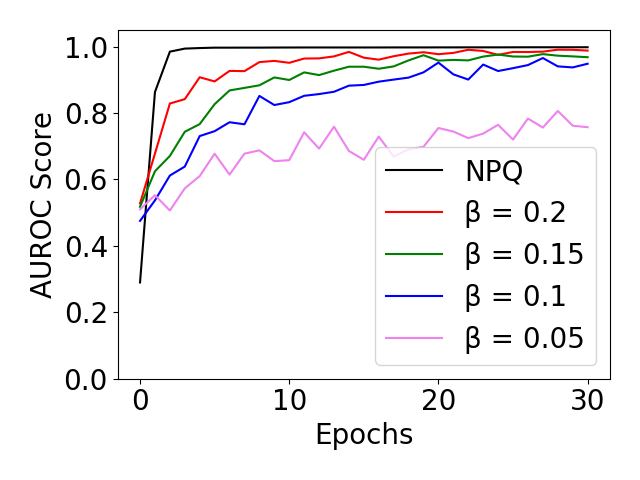}
  \caption{$5$ clients and $b = 16$.}
  \label{bc5.fig}
\end{subfigure}
\begin{subfigure}[t]{0.24\textwidth}
\centering
  \includegraphics[width=\textwidth] {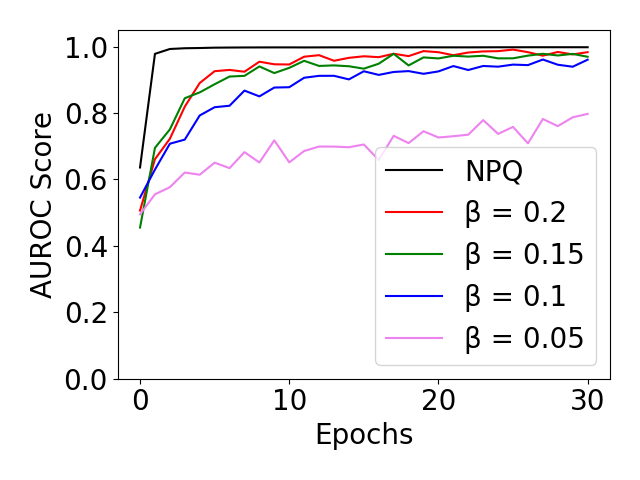}
  \caption{$10$ clients and $b = 16$.}
  \label{bc10.fig}
\end{subfigure}
\caption{AUROC Score on Breast Cancer dataset. NPQ (No Privacy noise and No Quantization) represents the baseline case without privacy noise and quantization.}
\label{bc.fig}
\vspace{-2ex}
\end{figure}

\section{Experiments}

\begin{table}[b]
\caption{Average aggregation time per epoch for Breast Cancer dataset with $\beta$=0.05 for the first $10$ epochs.}
\begin{center}
\begin{tabular}{| p{1.1cm} | p{0.4cm} | p{2.3cm} | p{2.3cm} |}
\hline
Number of clients & $b$ & Off-chain aggregation time (seconds) & On-chain aggregation time (seconds)\\
\hline\hline
  & 2 & 0.00023 & 12.090\\
5 & 4 & 0.00023 & 12.389\\
  & 8 & 0.00021 & 12.432\\
  & 16 & 0.00019 & 12.521\\
\hline
   & 2 & 0.00024 & 21.209\\
10 & 4 & 0.00022 & 21.721\\
   & 8 & 0.00021 & 21.727\\
   & 16 & 0.00021 & 21.723\\
\hline
\end{tabular}
\end{center}
\label{table:perepochtimes}
\end{table}

We present experiments to evaluate the performance of Algorithm~\ref{alg} on two different medical datasets, implemented as an off-chain standalone VFL application and an on-chain DP-BBVFL decentralized application.
Our code is available at \url{https://anonymous.4open.science/r/F23_HealthFederated-D196}.

\subsection{Datasets}

\noindent\textbf{Breast Cancer} \cite{misc_breast_cancer} is a small-scale dataset of $569$ samples with $30$ features for binary classification. We experiment with $5$ and $10$ clients, each holding $6$ or $3$ features. We use a three-layer neural network as the local model for each client and a fully connected layer for classification as the fusion model. We use batch size $B = 10$ and embedding size $P = 16$ and train for $30$ epochs with a learning rate $0.001$. We evaluate the accuracy of the model with the AUROC score.

\noindent\textbf{MIMIC-III} \cite{mimic-article} is a large-scale time-series dataset of $60,000$ intensive care unit admissions. We use a subset of the dataset for binary prediction of in-hospital mortality. The subset includes $8,629$ samples, each with $60$ time frames and $76$ features, including patient information such as demographics, vital signs, medications, etc. We experiment with $4$ clients, each holding $19$ out of $76$ features. Each client's local model is a two-layer LSTM, and the fusion model is a fully connected layer for binary classification. We use batch size $B = 100$ and embedding size $P = 16$ and train for $200$ epochs with a learning rate $0.0001$. We evaluate the accuracy of the model with an F1 score.

\subsection{Experiment Setup}

To test our setup, we used a private blockchain constructed using a Python-native Ethereum Virtual Machine (EVM), i.e., PyEVMBackend. Our blockchain was deployed on a system with a 12-core CPU and 36 GB RAM. 
To evaluate the performance of Algorithm~\ref{alg} with different levels of privacy guarantee, we ran the experiment with various PBM noise parameters. We used $b=\{2,4,8,16\}$ and $\beta=\{0.05, 0.1, 0.15, 0.2\}$ for the Breast Cancer dataset, and $b=\{8,16,64\}$ and $\beta=\{0.1, 0.15, 0.2\}$ for MIMIC-III. We compare the performance of DP-BBVFL with the baseline case (NPQ) where the model is trained without PBM privacy noise and quantization. We note that a direct comparison with other works on private VFL with or without blockchain is unworkable because those works either employ different privacy methods (e.g., functional encryption), use a less general model (e.g., a k-means algorithm), or have different privacy goal (e.g, intention-hiding).

\subsection{Results}

Fig.~\ref{bc.fig} shows the test AUROC score by epoch for the Breast Cancer dataset, and Fig.~\ref{mimic.fig} shows the test F1 score by epoch for MIMIC-III. 
In both Fig.~\ref{bc5.fig}  and Fig.~\ref{bc10.fig}, the baseline case performs the best, which is expected because there is no privacy noise. For a fixed value of $b$, if we lower the privacy level by increasing $\beta$, the accuracy increases. However, higher $\beta$ provides less privacy, so there is a tradeoff between accuracy and privacy. We observe the same trend in Fig.\ref{mimic16.fig} and Fig.\ref{mimic64.fig}.

\begin{figure}[t]
\centering
\begin{subfigure}[t]{0.24\textwidth}
\centering
  \includegraphics[width=\textwidth]{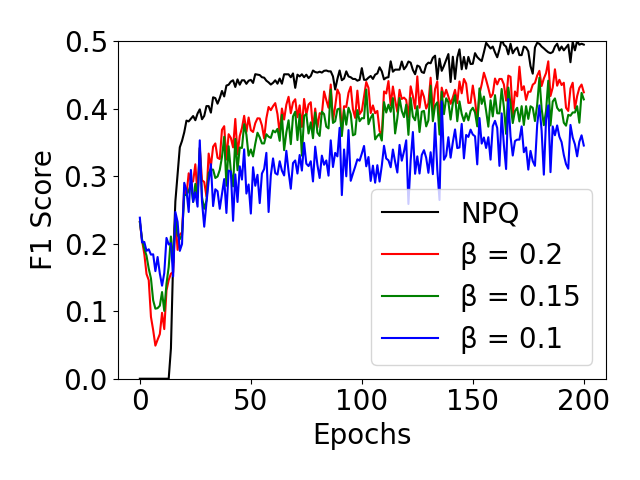}
  \caption{$4$ clients and $b = 16$.}
  \label{mimic16.fig}
\end{subfigure}
\begin{subfigure}[t]{0.24\textwidth}
\centering
  \includegraphics[width=\textwidth]{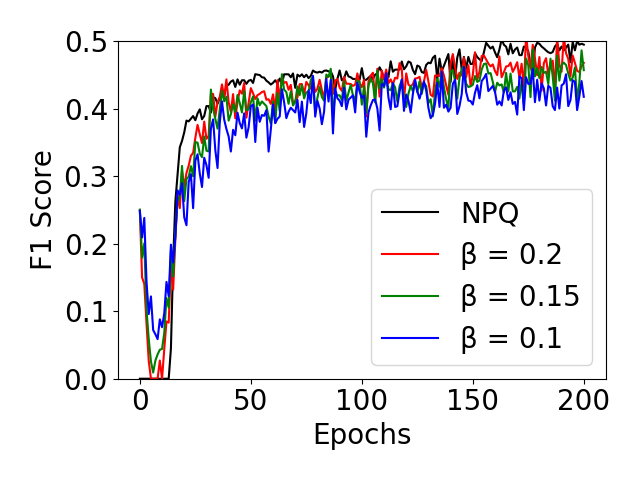}
  \caption{$4$ clients and $b = 64$.}
  \label{mimic64.fig}
\end{subfigure}
\caption{F1 Score on MIMIC-III dataset. NPQ (No Privacy noise and No Quantization) represents the baseline case without privacy noise and quantization.}
\label{mimic.fig}
\vspace{-2ex}
\end{figure}

Table~\ref{table:perepochtimes} shows the average time per epoch to aggregate the embeddings for the first $10$ epochs on-chain using the smart contract aggregator. It should be noted that the aggregation time on the blockchain is considerably slower than off-chain. This is expected as blockchain computation is generally time-consuming due to the inherent consensus mechanism associated with the transaction recording process. This is a trade-off for ensuring transparency and verifiability. In addition, our on-chain aggregator smart contract does not have optimizations for fast tensor processing that popular Python libraries like PyTorch that were used in the off-chain application. In addition to measuring on-chain aggregation time, in a large blockchain setup, throughput and latency would also be measured. As discussed in \cite{yin2019hotstuff}, these measurements can be performed using the ThroughputLatencyServer and ThroughputLatencyClient micro-benchmark programs \cite{bftsmart}.

\section{Conclusion}

In this work, we proposed DP-BBVFL, a blockchain-based serverless VFL algorithm designed for applications with vertically partitioned data to train a task collectively. We posit that traditional VFL has trust and privacy concerns due to its reliance on a central server. We leverage blockchain technology and DP in our algorithm to address these issues. Using blockchain promotes verifiability and eliminates the need for trust, and applying PBM protects the embeddings when exposed on-chain. Our algorithm allows clients to participate in a more transparent VFL task while preserving data privacy. 
Through rigorous experimentation with medical datasets for applications mimicking decentralized healthcare-focused scenarios, we demonstrate that DP-BBVFL achieves accuracy levels comparable to contemporary models. However, this comes with a tradeoff in training time due to the computational demands of on-chain aggregation. Despite this, the benefits of enhanced privacy and verifiability make DP-BBVFL a compelling solution for sensitive applications.
Optimizations for tensor processing in on-chain aggregation could considerably improve the aggregation time, and there is scope for future work in this area. In addition, there is scope for investigation into the effect of blockchain latency and throughput on the overall training time.
%
Overall, DP-BBVFL represents a groundbreaking step forward in the utilization of VFL in environments where data privacy and model verifiability are paramount with vast and varied decentralized application domains. 

\section{Resource Contributions}
Our code and documentation are shared under the Apache 2.0 license in an open-source GitHub repository: \url{https://github.com/AI-and-Blockchain/F23_HealthFederated}.

\nocite{*}
\bibliographystyle{abbrv}
\bibliography{references}

\end{document}